\documentstyle[twocolumn,aps,epsfig]{revtex}

\newcommand{\fig}[1]{Fig.~\ref{#1}}

\begin{document}
\title { 
        Search for pairing-vibration states of even Ca isotopes
        in $^{40}$Ca+$^{208}$Pb transfer reactions
       } 
\author{S. Szilner, L. Corradi, B.R. Behera, E. Fioretto, A. Gadea,
         A. Latina, A.M. Stefanini \\ M.Trotta,
         A.M. Vinodkumar, Y.W. Wu} 
\address
{Istituto Nazionale di Fisica Nucleare, Laboratori Nazionali di Legnaro,
I-35020 Legnaro, Italy}
\author{S.Beghini, G.Montagnoli, F.Scarlassara }
\address
 {Dipartimento di Fisica, Universit\'a di Padova, and Istituto Nazionale
 di Fisica Nucleare, I-35131 Padova, Italy}
\author{G.Pollarolo}
\address 
{ Dipartimento di Fisica Teorica, Universit\'a di Torino, and
 Istituto Nazionale di Fisica Nucleare, 10125 Torino, Italy}
\author{F. Haas, E. Caurier, F. Nowacki}
\address{Institut de Recherches Subatomiques, 
(IN2P3-CNRS-Universit\'{e} Louis Pasteur), 
 F-67037 Strasbourg Cedex 2, France}

\date{\today}
\maketitle

\begin{abstract}
 Multi-nucleon transfer reactions in $^{40}$Ca+$^{208}$Pb have been 
 studied at several bombarding energies close to the Coulomb barrier. 
 Light reaction products have been identified in mass and charge 
 with a time-of-flight spectrometer.    
 The energy spectra of the inclusive two neutron pick-up channel
 show a population in a narrow  
 region of excitation energies which corresponds to the 
 predicted energy of pairing vibration states in
 $^{42}$Ca.
\end{abstract}
%\maketitle
%  
% test start here
%
\section{Introduction}   
%%%%%%%%%%%%%%%%%%%%%%%%%%%%%%
In the nuclear medium, the ability of nucleons to form 
0$^+$ pairs has been realized 
very early from the study of nuclear masses, but the recognition that 
these pairs of nucleons play an important role in the description of 
nuclear spectra, 
in particular in correlating spectra of neighbouring nuclei,
is more recent \cite{bm,bohr,nathan}. 
The development of the pair model 
received considerable inputs from the extensive experimental work 
on transfer with light ions, in particular 
$(p,t)$ and $(t,p)$ reactions  \cite{broglia,light}.    
The use of heavy-ions was perceived \cite{dietrich} 
as a very promising tool for the 
study of 
pair rotation and/or pair vibration 
degrees of freedom since one could envisage the exchange 
of many nucleons among the reactants.  
Unfortunately, the lack of sufficient 
mass, charge and energy resolution postponed 
the experimental studies to recent times, 
when high resolution spectrometers 
became available and the clear identification of transfer products 
up to 
at least four neutron pick-up and 
four proton stripping has been achieved  
\cite{rehm,mntar2,wvo,ni58pb208}.    
In these measurements, attempts to identify 
pair transfer modes were based on total or
differential cross sections, but no unambiguous 
contribution of these degrees of freedom could be drawn. 

Systematic studies of $(p,t)$ and $(t,p)$ reactions 
led to the identification of the calcium region as the only known 
region where the 
cross sections for the population of the excited 0$^{+}$ states 
is larger than the ground state. 
Those states   
have been recognised as multi 
(addition and removal) pair-phonon states 
\cite{broglia}. 
Nuclear structure 
and reaction dynamics
studies attribute this behavior 
to the influence of the $p_{3/2}$ orbital that gives a much larger 
contribution to the two-nucleon transfer cross section than the 
$f_{7/2}$ orbital, which dominates the ground state wave function. 

In the present work we report on a study of multi-nucleon transfers 
in $^{40}$Ca+$^{208}$Pb. Since 
both nuclei are recognised as pair vibrational, they provide 
an ideal tool to study multi pair-phonon excitations 
by looking at $Q$-value 
distributions for the different isotopes. 
Since this system is closed shell, both in proton and neutron, it 
provides also an excellent opportunity for a quantitative comparison 
with a theoretical model whose results depend strongly on 
the form factors defining the transfer process \cite{abs1,winther-wkb}.

\section{Results and discussion}
%%%%%%%%%%%%%%%%%%%%%%%%%%%%%%%%%%%%%%%
The experiment was performed using the Tandem + ALPI accelerator 
of the Laboratori Nazionali di Legnaro. 
A $^{40}$Ca beam was accelerated onto a $^{208}$Pb target 
(200 $\mu$g/cm$^{2}$)
at several  bombarding energies close to the Coulomb barrier: 
$E_{\rm lab}$= 250, 236 and 225 MeV.  
Light reaction products were detected with the   
spectrometer PISOLO \cite{ni58pb208,pisolo}, that makes use of 
two micro-channel plate detectors for time-of-flight signals 
and a $\Delta E/E$ ionization chamber for  
nuclear charge and energy determination. 
The good mass and charge resolution of $\Delta A/A \simeq 1/120$ and
$\Delta Z/Z \simeq 1/60$, respectively, 
allowed a clear separation between different channels. 
Taking into account  
the intrinsic energy resolution of the detector and
the energy straggling in the target the obtained energy resolution was 
 $\simeq$ 2 MeV.  
With a total solid angle of $\simeq$ 3 msr, transfer cross sections 
up to the pick-up of seven neutrons and the stripping of 
seven protons have been extracted.
In the present paper we focus on pure neutron transfer channels. 

The comparison of the data is done with the Complex-WKB model 
\cite{winther-wkb,winther,s32pb208} that calculates the outcome 
of the reaction, namely total kinetic energy loss, 
differential and total cross sections, taking into account 
the transfer of single nucleons by using the well known one-particle 
transfer form factors. 
This model solves in an approximate way the system of 
coupled equations which determine 
the exchange of particles between the nuclei
by treating the particle
transfer degrees of freedom as independent and in the harmonic approximation.  

\begin{figure}
\begin{center}
\epsfig{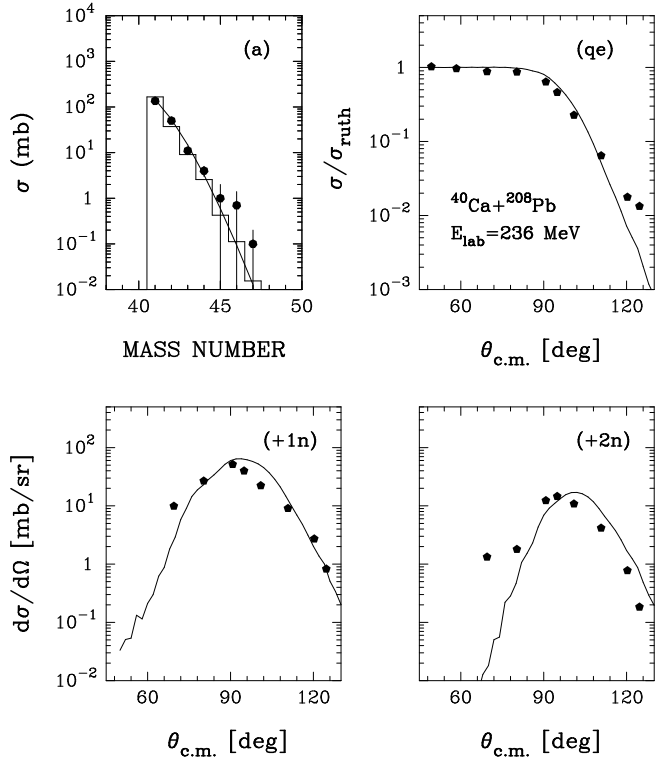}
\end{center}
 \caption{
  In (a) the angle and $Q$-value integrated cross sections 
  of the pure neutron pick-up
  channels for the $^{40}$Ca$+^{208}$Pb reaction at 
  $E_{\rm lab}=$236 MeV (dots) are shown as a function of 
  the mass number together with the
  theoretical (CWKB) prediction (histogram), 
  the curve corresponds to a Poisson distribution with
  an average number of transfered particles 
  $\langle n\rangle =1.1$.
  In the other frames the
  $Q$-value integrated angular distributions for the quasi-elastic (qe), 
  one-neutron (+1n) and two-neutron (+2n) pick-up channels 
  (dots) are shown together with the
  theoretical calculation (curves).
 }
\label{tcsang}
\end{figure}

\begin{figure}
\begin{center}
\epsfig{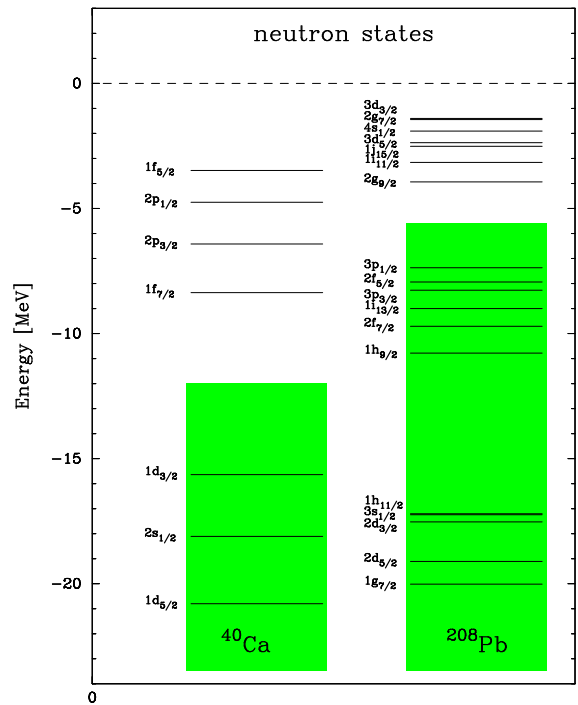}
\end{center}
 \caption{
  Neutron single-particle levels for projectile and target. 
  The shaded areas 
  indicate the occupied levels. The spectroscopic factors have been 
  chosen to be unity for all the levels.   
 }
\label{fig:spl}
\end{figure}

The experimental total cross sections and the angular distributions  
at $E_{\rm lab}=$ 236 MeV are shown in \fig{tcsang}    
together with the theoretical predictions.  
In order to cover the full range of transferred energies,   
the single-particle form factors governing 
the exchange of nucleons were 
calculated
by including
all single-particle levels above the Fermi surface and a full shell below. 
Those levels are shown in \fig{fig:spl} for neutrons. 
For the real part of the optical potential we used the empirical 
potential of Ref. \cite{aagebook}. 
For the imaginary part the model uses
only a volume term with a small diffusivity ({\em a} $\sim$ 0.4 fm) 
since the depopulation of the entrance channel is taken explicitly 
into account by the coupling to the transfer channels. 
This potential gives an adequate 
estimation of the total reaction cross section as can be inferred 
from the good description of the quasi-elastic angular distribution.     

The bell shaped distributions of the differential cross sections show 
the grazing character of the reaction. 
The integrated cross sections decrease smoothly as a 
function of the number of transferred neutrons, and the  
overall trend of the data is reproduced by calculations.  
We notice that the total cross section for the even calcium isotopes
is in agreement with a Poisson distribution, with
an average number of transferred neutrons $\langle n\rangle =$ 1.1,
that underlines the harmonic character of the
degrees of freedom determining the evolution of the reaction.
However from the observed behavior of the inclusive total cross sections 
alone, as was clear from previous experiments, %(you may put here refs)
it is difficult to draw unambiguous
conclusions on the role played by a pair transfer mode. 
As we will discuss below, the excitation functions, corroborated 
from the knowledge of the excitation spectra and from the results
of reaction and shell model calculations, 
provide valuable information in this respect. 

\begin{figure}
\begin{center}
\epsfig{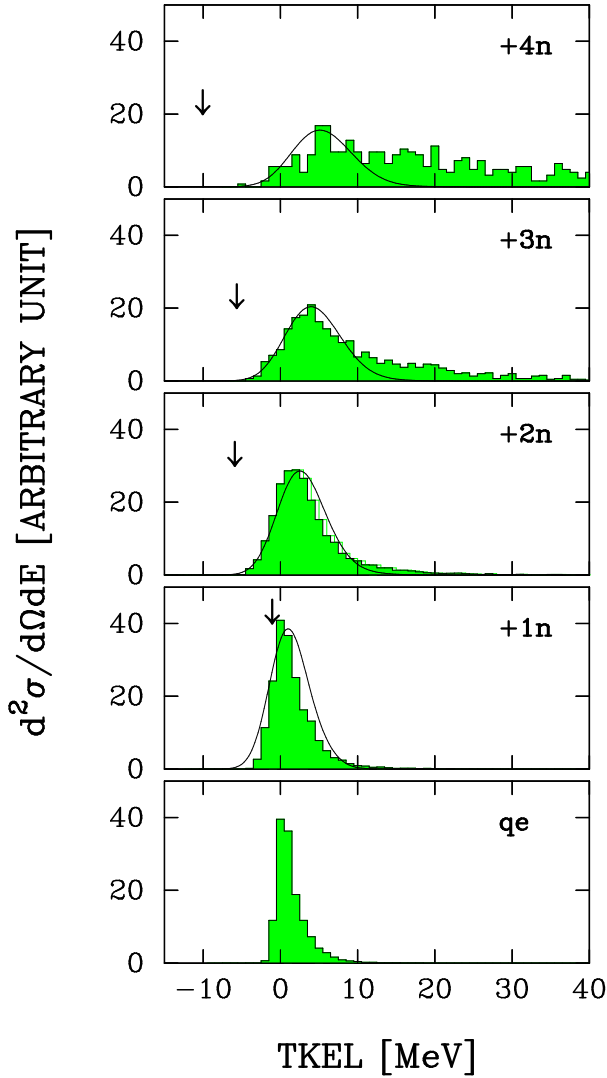}
\end{center}
 \caption{
  Total kinetic energy loss distribution of the quasi-elastic (qe) and
  neutron pick-up channels (histograms) together with the
  CWKB calculations (curves) measured at $E_{\rm lab}=$236 MeV.
  Arrows indicate the ground to ground state $Q$-value.
 The data and calculations have been normalised arbitrarily.
 }
\label{qvalue}
\end{figure}

For the bombarding energy of 236 MeV we show in \fig{qvalue} 
the total kinetic energy loss (TKEL) distributions measured at 
the scattering angle of $\theta_{\rm c.m.}=95^{\circ}$ for channels 
up to +4n in comparison with the theoretical prediction calculated 
for a partial wave close to the grazing one. 
In the figure we indicate with an arrow the ground to ground state 
$Q$-value. 
As can be appreciated,  
the multi-nucleon transfer  
channels (we recall that the optimum $Q$-value for all neutron transfer 
channels is close to 0)  
display a well defined maximum that is shifted to high energy losses, 
leaving the ground states unpopulated. 

\begin{figure}
\begin{center}
\epsfig{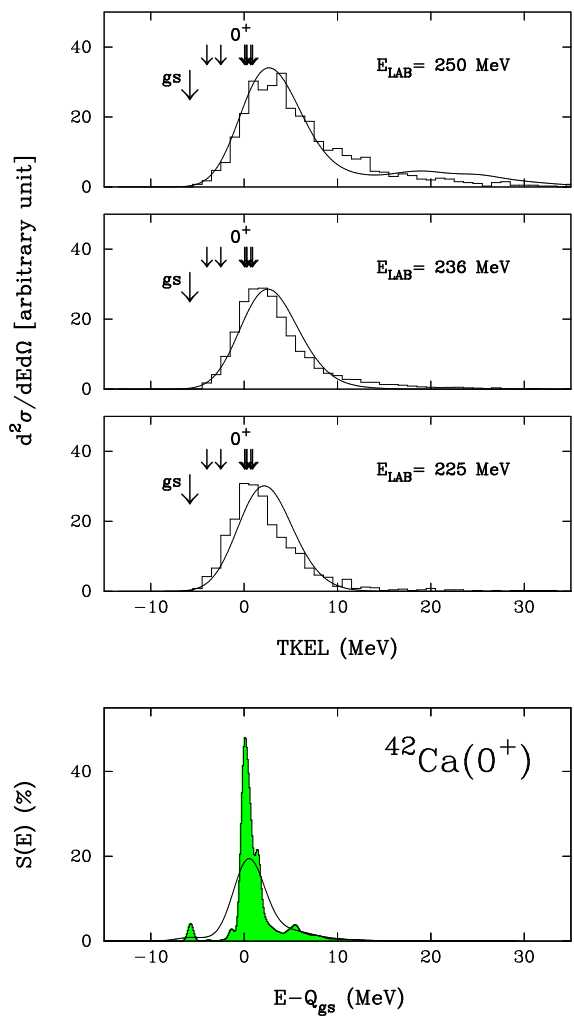}
\end{center}
\caption{
 Experimental (histograms) and theoretical (curves)
 total kinetic energy loss distributions of the
 two neutron pick-up channels at the indicated energies.
 The arrows correspond to the energies of 0$^+$ states in $^{42}$Ca
 with an excitation energy lower than 7 MeV [18].
 The group of states at 5.87 (collecting most of the strength),
 6.02, 6.51 and 6.70 MeV were strongly populated in $(t,p)$ reactions
 [20-22].
 The same excitation energy range was strongly populated in the
 $^{40}$Ca($^{14}$C,$^{12}$C)$^{42}$Ca reaction [23].
 Bottom panel shows the
 strength function $S(E)$  from SM calculations (see text)
 after convoluting
 with Gaussian of two different widths: 300 keV
 and 1.5 MeV (close to the experimental energy resolution - curve).
 The represented strength function has been obtained after 200 Lanczos
 iterations to allow a correct convergence of all eigenstates.
}
\label{qvalue-2n}
\end{figure}

The calculation 
gives an overall good description of the experimental spectra 
underestimating the high energy tails especially for the massive transfer 
channels.
The discrepancy may be partially overcome by 
better treatment of small impact parameters 
{\it i.e.} deeper penetrating trajectories, 
that gradually become 
more important as more particles are transferred.  
The theory calculates the redistribution of mass 
and charge by using an independent particle description thus the 
calculated spectra reflect the underlying single particle level structure.   
By looking at the final population of the single particle levels
we can infer that the maxima for the +2n and +4n channels
are essentially due to two and four neutrons in the $p_{3/2}$ orbital.
This fact together with the known low energy spectra of $^{42,44}$Ca
\cite{ca42level} 
suggest that these maxima correspond to the excited $0^+$ 
states that were  identified with the 
pair mode \cite{broglia,light}.
Using $^{48}$Ca as the ground state of the pair
vibrational model these states are interpreted as the mode 
$(4,1)$ in $^{42}$Ca and $(4,2)$ in $^{44}$Ca where
the first number in parentheses refers to the
pair-removal $(n_r)$ and the second to the pair addition $(n_a)$ mode.
We recall that the pair vibrational model assigns the energies
with the formula $E(n_r,n_a)=\hbar\omega_p(n_r+n_a)$, 
 $\omega_p$ being the
pair frequency. In the case of calcium $\hbar\omega_p$ is $\simeq$ 3 MeV.
To strengthen this assignment, for the +2n
channel we show in \fig{qvalue-2n} the spectra taken at the three bombarding
energies; here arrows label the position of the 0$^+$ states.

The indicated maxima of the spectra are compatible with
the excitation of $0^+$ states in $^{42}$Ca at $\simeq$ 6 MeV, 
leaving room for the mutual excitation
of the yrast states in $^{206}$Pb \cite{lanford}.
We notice here that the spectra widen with the increase of the
bombarding energy with a slight shift of the maxima reflecting
the energy dependence of the optimum $Q$-value.
Both features are well reproduced by the theoretical calculations
indicating that the used single particle levels cover the full $Q$-value
ranges spanned by the reaction.

 Since the illustrated experimental results and reaction calculations
 point to a 
 selective feeding of $0^{+}$ states dominated
 by a pair of neutrons in the $p_{3/2}$ orbital 
 we felt that it would be interesting to calculate, for $^{42}$Ca,
 the strength distribution of 
 these peculiar $0^{+}$ states over all others $0^{+}$ states. 
 This has been performed in the framework 
 of large scale shell model (SM) calculations by using 
 the same model space and interaction as in a 
 recent 
 publication concerning various spectroscopic 
 features of calcium isotopes \cite{sm1}. 
 To extract the channel strength function we used the Lanczos method  
 by using as pivot state  
 $\psi_{1} = (a^{+}_{p_{3/2}} a^{+}_{p_{3/2}})^{0}|0^{+}\rangle$
 that corresponds to the creation of two neutrons, coupled to 0,  
 on the $|0^{+}\rangle$ ground state of $^{40}$Ca.
 The valence space, used in these SM calculations
 consists of a $^{28}$Si inert core and of 
 the 2$s_{1/2}$, 1$d_{3/2}$, 1$f_{7/2}$ and 2$p_{3/2}$ 
 subshells for both protons and neutrons. 
 The ground state of $^{40}$Ca is thus described by a 
 $(np-nh)$ configuration with $n$ up to 12.
 Through the Lanczos iterative procedure, and using the overlap 
 relation between the obtained eigenstates 
 of the SM Hamiltonian  and the pivot state $\psi_{1}$,  
 we can calculate \cite{sm2,sm3}  
 the strength function for these peculiar $0^+$ states.
 The strength distribution $S(E)$,
 shown in the lower panel of Fig. \ref{qvalue-2n},
 displays, clearly, a strong concentration 
 near $\sim 6$ MeV of excitation energy,  
 an energy very close to that of a configuration 
 where a $p_{3/2}$ neutron pair is coupled to a 
 closed shell of $^{40}$Ca ground state ($E \sim 5.9$ MeV). 
 This calculation demonstrates the dominant 
 contribution of the $p_{3/2}$ orbitals
 and the predicted very narrow energy distribution suggests its 
 interpretation as a pair mode.

\section{Summary}
Multi-neutron transfer reactions, 
at several bombarding energies, have been studied in 
$^{40}$Ca+$^{208}$Pb.  
Due to the peculiar properties of 
the $f$-$p$ shell, this system allows us to interpret the total 
kinetic energy spectra as being dominated by the excitation of 
pair modes in even calcium isotopes. 
The population of these channels is in agreement with 
the harmonic properties of these modes. 
This experiment demonstrates 
the importance of pursuing these studies with heavy ions even if, 
within the present energy resolution, a  
definite assignment of these states is not possible. 
The large measured cross sections  would allow 
the interesting study of the decaying pattern of these pair modes.
  
%%%%%%%%%%%%%%%%%%%%%%%%%%%%%%%%%%%%%%%%%%%%

%%%%%%%%%%%%%%%%%%%%%%%%%%%%%%%%%%%%%%%%%%
\end{document}